# Sustaining Efficiency at Elevated Power Densities in InGaAs Airbridge Thermophotovoltaic Cells

Bosun Roy-Layinde[1], Tobias Burger[1], Dejiu Fan[2], Byungjun Lee[2], Sean McSherry[1], Stephen R. Forrest[2,3,4], Andrej Lenert[1]

[1]Department of Chemical Engineering, University of Michigan, Ann Arbor, MI, 48109

[2]Department of Electrical Engineering and Computer Science, University of Michigan, Ann Arbor, MI, 48109

[3]Department of Physics, University of Michigan, Ann Arbor, MI, 48109

[4]Department of Materials Science and Engineering, University of Michigan, Ann Arbor, MI, 48109

E-mail: alenert@umich.edu;

## ABSTRACT

Here we investigate the use of single-junction InGaAs airbridge cells (ABCs) at elevated power densities. Such conditions are relevant to many thermophotovoltaic (TPV) applications, ranging from space to on-demand renewable electricity, and require effective management of heat and charge carriers. Experimental characterization of an InGaAs ABC with varying emitter and cell temperature is used to develop a predictive device model where carrier lifetimes and series resistances are the sole fitting parameters. The utility of this model is demonstrated through its use in identifying near-term opportunities for improving performance at elevated power densities, and for designing a thermal management strategy that maximizes overall power output. This model shows that an InGaAs ABC with material quality that leads to the longest reported carrier lifetimes can attain efficiencies exceeding 40% at 0.5 W/cm$^2$, even when considering the power necessary to cool the cells.

# I. INTRODUCTION

Thermophotovoltaic (TPV) systems offer a solid-state approach for converting heat into electricity with various potential advantages over conventional heat engines. These include scale-insensitive performance above ~10 kW [1] which is particularly desirable in applications such as remote power generation and co-generation of heat and power near the point of use [2–5]. Further, TPVs are capable of near -ramping which is important for regulating the supply of intermittent renewable energy sources like wind and solar [3–5].

The airbridge cell (ABC) architecture is a promising approach to improve the efficiency of TPV systems by enabling near-perfect reflection of out-of-band (OB) thermal radiation, thereby overcoming the performance of prior spectral control strategies [9–18] [19]. Specifically, the addition of a ~600-nm deep air pocket below an $In_{0.53}Ga_{0.47}As$/InP heterojunction improves the efficiency of converting absorbed thermal radiation into electrical power from approximately 23% to 32% [10]. The large refractive index contrast between the air gap and the surrounding layers allows the ABC to reflect ~99% of OB power emitted by a 1182°C SiC emitter.

Despite the potential advantages offered by the ABC, the performance of InGaAs ABCs currently peaks at relatively low power densities (≤0.3 W/cm$^2$) and does not account for power losses for thermal management of the cell. Sustaining high efficiency at elevated power densities is important for practical deployment of thermophotovoltaic (TPV) applications, including grid-scale energy storage [5,7,8,21,22] and distributed generation [3,4,23]. In particular, TPV cells are likely to see elevated emitter temperatures, and therefore elevated power densities, in energy storage applications to maximize their round-trip efficiency [7,8]. Furthermore, effective thermal management of the cells becomes an integral aspect of the overall design under such conditions.

Here we investigate how elevated power densities affect the performance of InGaAs ABCs and propose strategies to overcome existing limitations. The study consists of experiments and models that enable design. We experimentally measure the optical and current-voltage characteristics of the InGaAs ABC while varying the cell temperature from 20°C to 75°C, and the SiC emitter temperature from 804°C to 1177°C. These data are used to parametrize the device model that captures various temperature-related effects, including bandgap narrowing and increased recombination. The model, in turn, allows us to analyze how increasing the power density affects the performance of the three relevant cell designs shown in Fig 1: (a) a cell with a planar metallic back surface reflector, designated as 'planar', with an OB reflectance of ~95%, (b) the previously demonstrated InGaAs ABC [20], and (c) a closer to ideal ABC (termed ABC*) with improved carrier management, corresponding to the longest reported lifetimes in InGaAs [24]. We also consider the power required to cool the cells. These considerations reveal that overall efficiencies exceeding 40% can be achieved at higher power densities (0.5 W/cm$^2$). Notably, this operating point corresponds to an emitter temperature of ~1070ºC, which is relatively low and consequently may enable the deployment of ABCs in applications such as waste heat harvesting and solar thermal power generation. Overall, this study addresses important questions related to using ABCs in energy systems and provides a near-term pathway to achieving high performance.

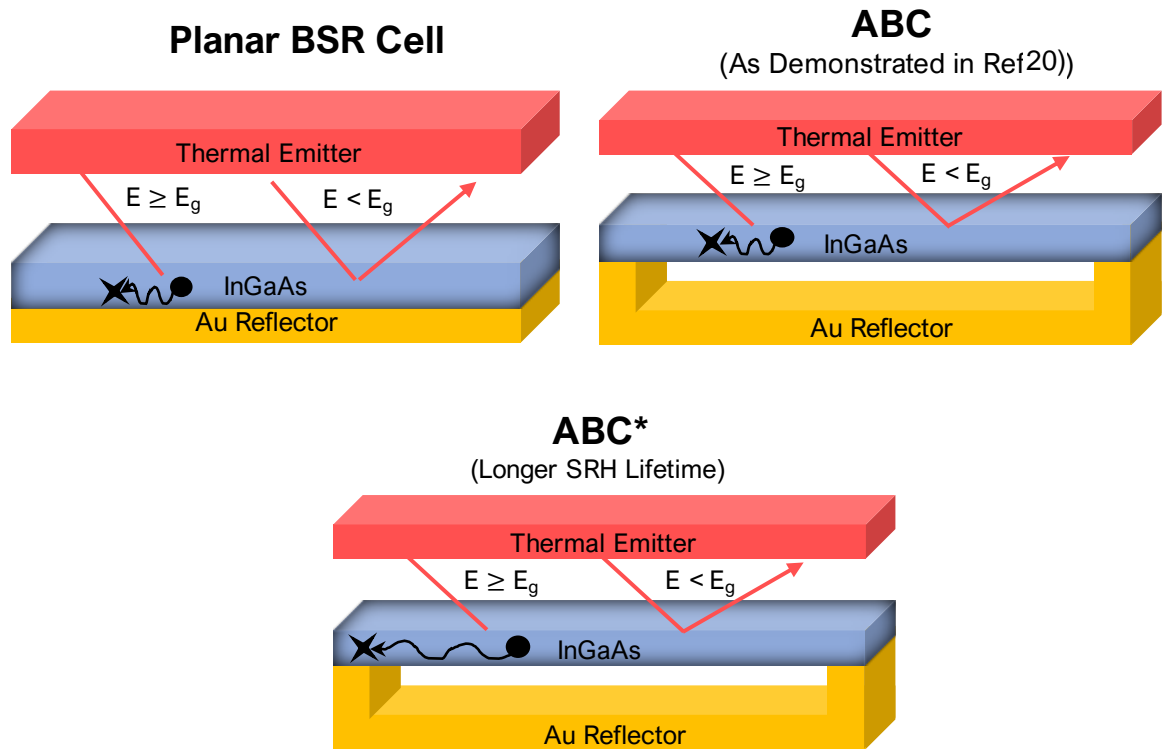

**FIG. 1 | Three TPV configurations at the focus of this study**: **(a) Planar**: an InGaAs cell with a planar back mirror exhibiting out-of-band reflectance ($R_{out}$) of ~95%, **(b) ABC**: the InGaAs airbridge cell (ABC) demonstrated in Ref. [20] exhibiting close to 99% $R_{out}$, and **(c) ABC***: an ABC with better material quality (corresponding to longest reported lifetimes [24]).

## II. EXPERIMENT AND THEORY

### A. TPV characterization

Figures 2a and 2b show the experimental setup used for current density-voltage (*J-V*) characterization with variable emitter and cell temperatures. *J-V* characteristics are measured using a Keithley 2401 source meter. Emitter and cell temperatures are independently varied in our experiments. Specifically, emitter temperature is regulated from 804°C to 1177°C by controlling the electrical input power to the lamp, while the cell temperature is maintained at 20°C using a chilled water loop. The cell temperature is varied from 20°C to 75°C using the water chiller/heater, while the emitter temperature is maintained at 1159°C. The temperature of the emitter is determined through Fourier transform infrared spectroscopy (FTIR) characterization and fitting of the emission spectrum using our previous procedure [20]. Cell parameters, including dark currents and series resistances, are extracted from *J-V* measurement (see Sec. II.B).

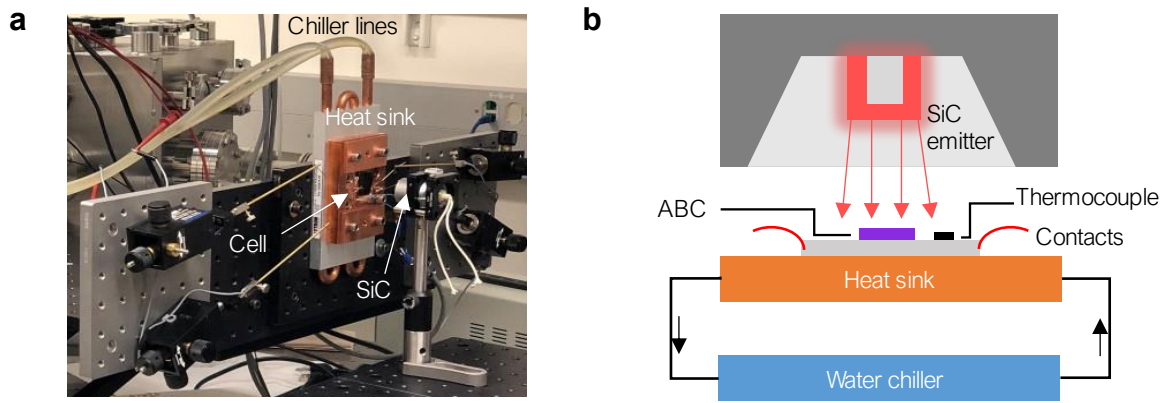

**FIG. 2 | Experimental characterization of temperature dependence. a)** Image and **b)** schematic of the temperature dependence measurement setup. A SiC globar emitter is used as the source of thermal radiation. The water chiller is used to cool and regulate the cell's temperature.

In addition to dark and illuminated *J-V* characterization with varying cell temperature, spectral reflectance of the ABC versus cell temperature is characterized at an incidence angle of 30° over a spectral region from 0.3 eV to 1.35 eV using an Agilent Cary 620 FTIR microscope and a heated stage. Absorption data are supplemented outside the specified range (< 0.3 eV and > 1.35 eV) by simulated results based on transfer matrix modeling [20].

## B. Device Model

The power generated by a TPV cell is equal to the product, $JV$. $J$ is equal to the photogenerated current density ($J_{ph}$) minus the current lost to radiative ($J_{rad}$) and non-radiative recombination ($J_{nrad}$) according to:

$$J = J_{ph} - (J_{rad} + J_{nrad}) \tag{1}$$

Shunt currents are neglected because they are small relative to the overall current density under normal illumination conditions. The other relevant current densities are described as follows:

(1) **Photogeneration**. $J_{ph}$ is the product of the incident photon flux $b(E, T_h)$, apparent view factor $F_v$, absorptance of the cell $a(E)$, and the internal quantum efficiency $IQE(E)$ according to:

$$J_{ph} = F_v \cdot q \int_{E_g}^{\infty} a(E) \cdot b(E, T_h) \cdot IQE(E) \, dE, \tag{2}$$

where the photon flux is given by Planck's law:

$$b(E, T_h) = \frac{2\pi E^2}{c^2 h^3 \left(\exp\left(\frac{E}{k_B T_h}\right) - 1\right)}. \tag{3}$$

Here, $E$ is photon energy, $T_h$ is the emitter temperature, $q$ is the electron charge, $c$ is the speed of light, $h$ is Planck's constant, and $k_B$ is the Boltzmann constant. The absorptance of the cell $a(E)$ is measured and compared to calculations using transfer matrix methods.

(2) **Non-radiative recombination.** The non-radiative recombination current density is given by:

$$J_{nrad} = J_{O1}\left(e^{\frac{qV'}{k_B T_c}} - 1\right) + J_{O2}\left(e^{\frac{qV'}{2k_B T_c}} - 1\right) + J_{Aug} \tag{4}$$

where $J_{O1}$ and $J_{O2}$ denote the Shockley-Read-Hall (SRH) saturation current densities of the charge carriers in the bulk (quasi-neutral) and depletion (space charge) regions of the cell, respectively, and $J_{Aug}$ is the Auger recombination current. $V' = V - JR_s$ accounts for the series resistance $R_s$. Consistent with prior modeling of thin heterojunctions with an n-type InGaAs absorber [25–28], $J_{O1}$ and $J_{O2}$ are approximated as follows:

$$J_{01} \approx q n_i^2 \left( \frac{1}{N_D} \sqrt{\frac{D_p}{\tau_1}} \right) \tag{5}$$

$$\text{and} \quad J_{02} \approx \frac{q n_i W}{\tau_2}, \tag{6}$$

where $n_i$ is the intrinsic carrier concentration, $N_D$ is the concentration of donors, $\tau_1$ is the lifetime of minority holes in the quasi-neutral region, $D_p$ is the diffusion coefficient of minority holes, $W$ is the depletion width, and $\tau_2$ is the lifetime in the depletion region. This model uses literature data for modeling the temperature dependence of the bandgap [25]. The dependence of the intrinsic carrier concentration on cell temperature is given by:

$$n_i = \sqrt{N_c \cdot N_v} \exp\left(\frac{-E_g}{2 k_B T_c}\right) \tag{7}$$

where $N_c$ ($N_v$) is the effective density of states in the conduction (valence) band.

We note that $\tau_1, \tau_2$ and $R_s$ are parameters specific to the ABC and are determined by fitting to experiments as discussed in II.C. Further, the temperature dependencies of $\tau_1, \tau_2$ and $R_s$ are assumed to be negligible.

The Auger recombination current is calculated by:

$$J_{Aug} = L(C_n + C_p) n_i^3 \exp\left(\frac{3qV}{2 k_B T_c}\right) \tag{8}$$

where $L$ is the thickness of the active region and $C_n$ and $C_p$ are the Auger recombination coefficients for recombination involving two holes and two electrons, respectively. For InGaAs, $C_n = C_p = 8.1 \times 10^{-29} \text{cm}^{-3}$ [24].

(3) **Radiative recombination**. Current resulting from radiative recombination is given by:

$$J_{rad} = J_{int} + J_{ext} \tag{9}$$

where $J_{int}$ describes photons that are parasitically absorbed within the cell by non-luminescent layers (i.e., no quasi-Fermi level splitting), while $J_{ext}$ describes the photons emitted by the cell. $J_{int}$ is calculated using the Multilayer Electromagnetic Solver for Heat Transfer [30] following previously reported procedures [20]. $J_{ext}$ is approximated by:

$$J_{ext} = qe^{\frac{qV}{k_B T_c}} \int_{E_g}^{\infty} a(E) \cdot b(E, T_c) \cdot IQE(E) \, dE \tag{10}$$

Here, we have used a Boltzmann approximation to describe the voltage-dependent emission. This model allows us to describe the effects of temperature on the *J-V* characteristics and other figures of merit of the TPV.

## C. Model Validation

To determine $\tau_1, \tau_2$ and $R_s$, the model is fit to (i.e., trained on) illuminated *J-V* measurements at a specific operation condition ($T_h$ = 1182°C, $T_c$ = 50°C) following a previous procedure [20], resulting in $\tau_1$ =13 ns, $\tau_2$ = 70 ns and $R_s$ = 30 mΩ.cm². The carrier lifetimes are consistent with previous studies [11], although they fall short of the best-reported values of 47 μs [11]. The accuracy of the model is then evaluated by comparing predicted *J-V* to measured data across the full range of experimental conditions. Experimental *J-V* characteristics of the ABC are shown alongside the model at varying cell temperatures in Fig. 3a. The root mean square error (RMSE) of the predictions for open-circuit voltage ($V_{oc}$), maximum power density ($P_{mpp}$) and short-circuit current density ($J_{sc}$) with varying cell temperature are 1.5%, 4.8% and 4.2%, respectively. Fig. 3b shows *J-V* data at three emitter temperatures while maintaining the cell at 20°C. The model predicts $V_{oc}$, $P_{mpp}$ and $J_{sc}$ with varying emitter temperatures within 0.65%, 1.7% and 0.08% RMSE, respectively. Overall, the error is less than 5% suggesting that the assumptions of temperature-independent carrier lifetime and series resistance are reasonable.

A decrease in the open-circuit voltage is observed with increasing cell temperature due to increased recombination. The decline in voltage is largely explained by the increasing intrinsic carrier concentration, which increases the rate of non-radiative recombination in the cell. In addition to lower voltage, we observe an increase in photocurrent density with temperature due to bandgap narrowing, as shown in Fig. 3c. The first absorption peak at energy higher than the bandgap, which is associated with an optical cavity resonance (within the cell) centered around ~0.73 eV, increases because of bandgap narrowing. We also observe a redshift (~2.14 ×10⁻² %/K) of the optical cavity resonance, consistent with a change in the refractive index of GaAs and InAs with temperature coefficients on the order of $10^{-2}$ %/K [31,32].

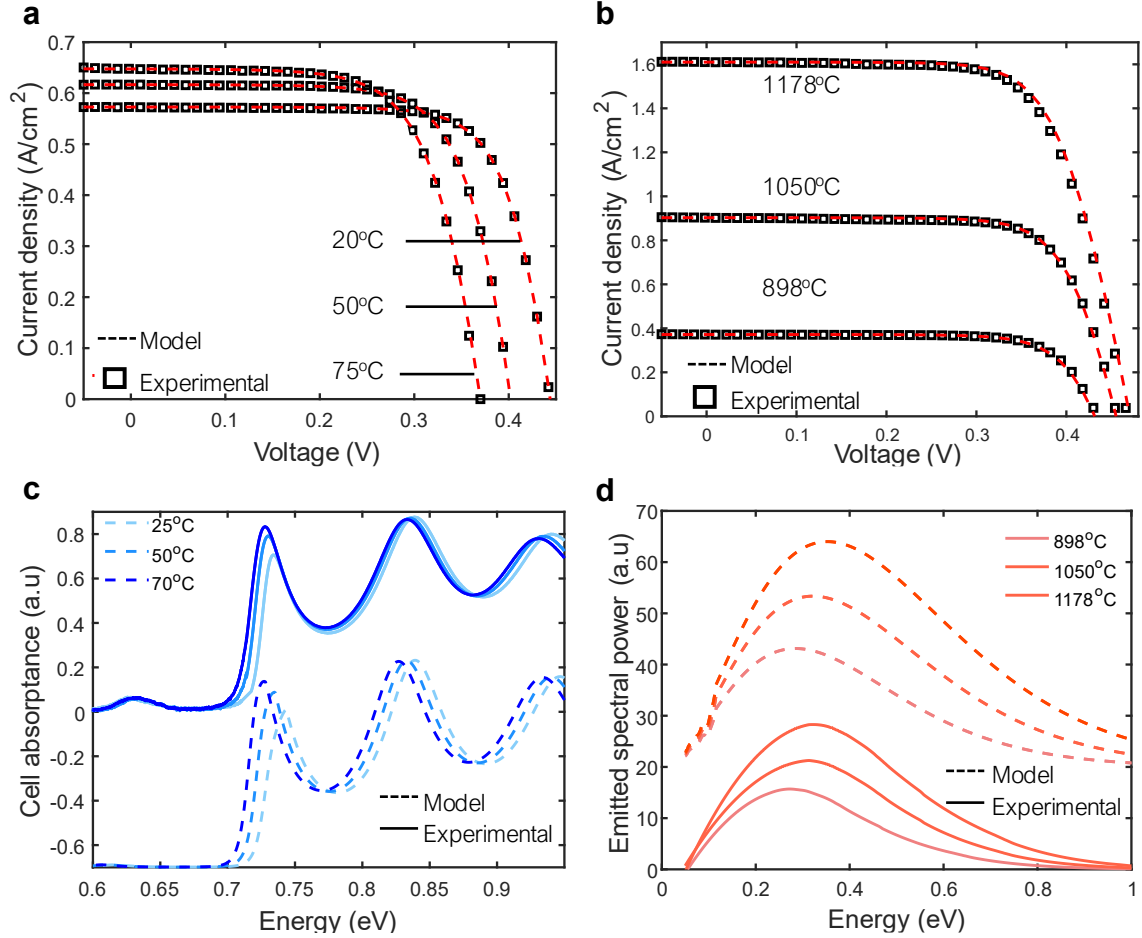

**FIG. 3 | Model training and validation. a)** Current density (*J*) vs voltage (*V*) curves at three representative ABC temperatures (20°C, 45°C, 75°C) and a constant emitter temperature of 1159°C. **b)** ABC *J-V* data at three representative emitter temperatures (898°C, 1050°C, 1177°C) and a constant cell temperature of 20°C. **c)** Spectral absorptance of the ABC at three representative cell temperatures. **d)** Spectral emission from the SiC emitter at three representative emitter temperatures. Experimental measurements: squares in a-b, solid lines in c-d; model predictions: dashed.

With increasing emitter temperature, there is an increase in the photocurrent density as more in-band photons are emitted (Fig. 3d). This effect is described by Planck's equation and the experimental emissivity of the thermal emitter. Accompanying this is a logarithmic rise in open-circuit voltage, inferred from Eqs. 1-3.

## D. Figures of Merit

The figure of merit used here is the absorption efficiency, defined as the ratio of generated power density ($P_{mpp}$) to the radiative heat flux absorbed by the cell ($Q_{abs}$) [8]. Absorption efficiency may be further decoupled into metrics specific to the quality of spectral and charge carrier management according to [8]:

$$\eta = \frac{P_{mpp}}{Q_{abs}} = (SE \cdot IQE)(VF)(FF) = \left(\frac{J_{sc}V_g}{Q_{abs}}\right)\left(\frac{V_{oc}}{V_g}\right)\left(\frac{P_{mpp}}{J_{sc}V_{oc}}\right). \tag{11}$$

Here, spectral management is defined by the product of spectral efficiency ($SE$) and the internal quantum efficiency ($IQE$); it describes how well the absorbed power is converted into the short circuit current multiplied by the bandgap voltage ($V_g$). Charge carrier management is described by the voltage factor ($VF$) and the fill factor ($FF$). Here, the voltage factor is the ratio of the open-circuit voltage to the bandgap voltage. The fill factor (last term on right in Eq. 11) largely reflects loss to series resistance and shunting [16].

## III. UNDERSTANDING AND OVERCOMING PERFORMANCE LOSS AT HIGH POWER DENSITIES

### A. Spectral and Carrier Management

Using the validated model described above, we focus on understanding how the three different systems behave at high power densities. Figure 4a shows the predicted efficiency versus power density of the InGaAs ABC compared to the planar cell and the higher performance ABC*. Even though ABC* can achieve better efficiencies at higher power densities relative to the others, we see a similar trend where the efficiency plateaus and begins to drop with increasing the power density.

To better understand the limitations of the existing ABC, the model is used to provide a breakdown of the loss pathways following photon absorption as a function of emitter temperature. As shown in Fig. 4b, non-radiative recombination is the dominant loss pathway, accounting for 36% of the power loss. We find that non-radiative recombination rate is largely controlled by defect mediated (SRH) recombination, and only <1% by Auger recombination. The other loss pathways include absorption of OB photons and thermalization of high energy photons. Although their overall fraction remains relatively constant at ~30% of the total, thermalization overtakes OB-loss as emitter temperature increases due to the blue shift in the blackbody distribution. Ohmic losses also constitute an important pathway for energy loss, particularly at high emitter temperatures because they scale quadratically with current density. We note this analysis is performed for a constant $F_v = 0.75$ (c.f. Eq. 2), and that operation at higher view factors will increase the photocurrent density produced and lead to higher Ohmic losses.

The above losses can be categorized as either affecting spectral management (e.g., OB loss and thermalization) or carrier management (e.g., recombination and Ohmic losses). The effects of increasing emitter temperature on spectral and carrier management of the ABC are depicted by the black curve in Fig. 4c. Remarkably, an increase of the emitter temperature from 667 to 1427°C changes the spectral management efficiency of the ABC by only ~15%, from 58 to 73%. The relative insensitivity of the ABC to emitter temperature is due to its very high reflectance (~98.5%) which limits OB loss from becoming dominant. In contrast, the spectral efficiency of the planar cell changes from 39% to 65% over the same emitter temperature range. The flattening of the ABC curve after reaching peak efficiency at ~1080°C indicates that a further emitter temperature increase has small effects on spectral management. Carrier management metrics, in contrast, are observed to sharply drop with increasing $T_h$ beyond that point, largely due to Ohmic losses.

To describe potential near-term performance improvements that may be achieved through the use of improved quality material, we model the performance of the ABC* cell, which exhibits the spectral properties of the ABC, but with the longest carrier reported for InGaAs (47μs) [24]. This results in an increased $V_{oc}$, increasing absolute efficiency over the entire emitter temperature range. Further, the carrier management metrics of the ABC* exhibit relative insensitivity to changes in emitter temperature, as Ohmic losses represent a smaller fraction of the output voltage. Notably, the efficiency of the ABC* peaks above 40% with an emitter temperature of ~1050°C.

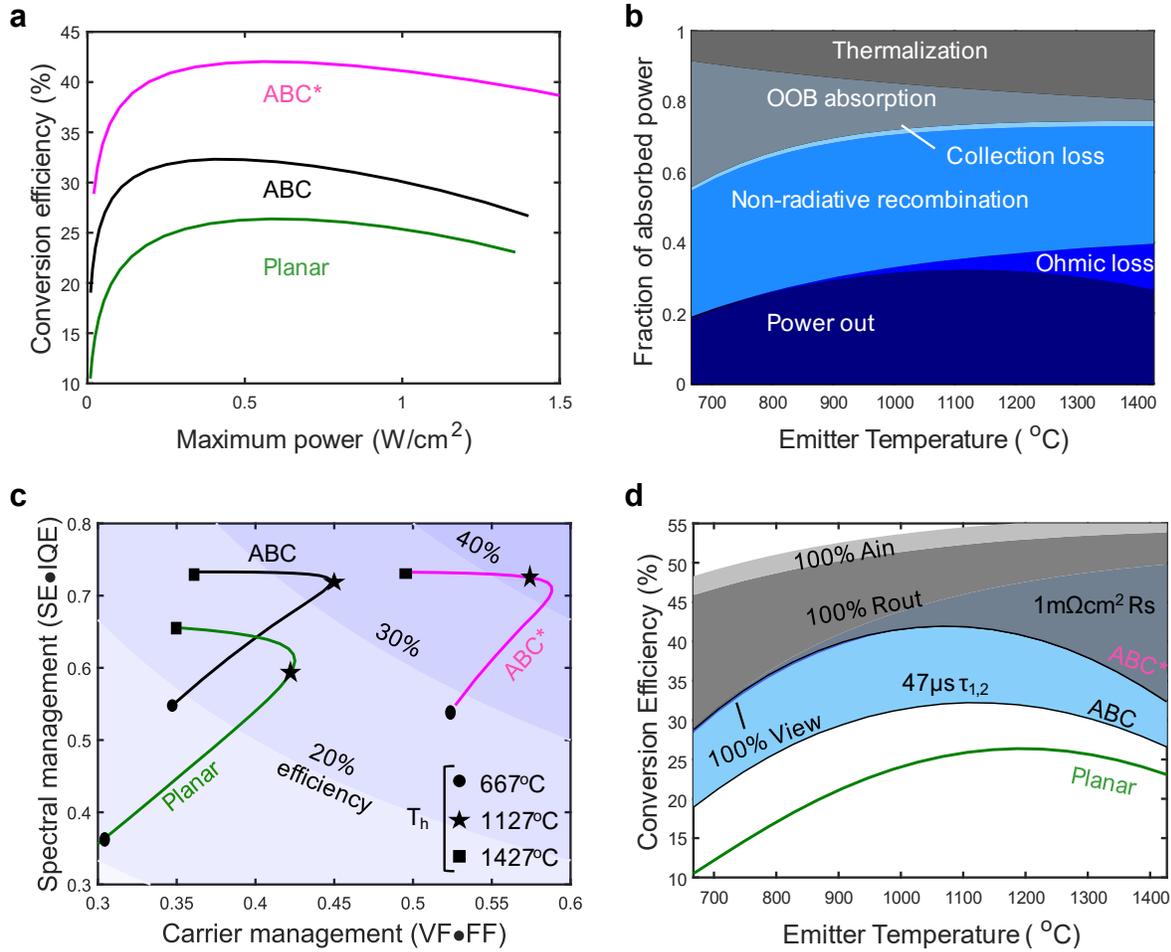

**FIG. 4 | Current ABC performance and near-term improvements. (a)** Predicted efficiency versus power for ABC, ABC* and planar configurations. **(b)** Loss pathways in ABC as a function of emitter temperature. **(c)** Efficiency plots for variable emitter temperature as a function of carrier and spectral management. Black, pink, and green solid curves represent simulated efficiency for the ABC, ABC*, and planar cell, respectively. **d)** Efficiency projection with improved spectral and carrier management.

To reach efficiencies exceeding 50%, additional advances are necessary, including reducing series resistance, increasing OB reflectance and IB absorption. Figure 4d shows a projection of conversion efficiency for various cell improvements as a function of variable $T_h$. Moving from ABC to ABC* is considered a near-term improvement, which yields a ~10% absolute increase in efficiency. Design of a TPV cavity with emitter-cell view factor approaching unity maximizes cell illumination and photocurrent, further increasing efficiency by ~0.6%. Beyond near-term improvements [34], decreasing cell series resistance from $30 m\Omega \cdot cm^2$ to $1 m\Omega \cdot cm^2$ would increase the conversion efficiency by ~5% absolute at 1127°C and enable operation at higher emitter temperatures and power densities. Optimized grids and selective contacts [35] may yield such reductions. Another ~5-15% absolute efficiency gain is observed in the limit of boosting the OB reflectance from 98.5% to 100%. Lastly, improving the absorption of in-band photons from 62% to ~100% using anti-reflective coatings and/or textured surfaces increases the output power, enabling another ~2% absolute efficiency increase. Overall, these cell improvements would enable >50% conversion efficiency at moderate emitter temperatures (i.e., <1000°C), while also sustaining high efficiency at the elevated power densities associated with high emitter temperatures.

## B. Thermal management

Active thermal management is needed at high power densities to maximize overall power output and mitigate temperature-related degradation of device lifetime [36–38]. An understanding of how heating affects efficiency allows us to calculate the heat load for each TPV system, and in turn, predict the amount of power consumed by the cooling system. The understanding is provided by the experimentally validated model that captures the effects of cell temperature on various optical and carrier-related mechanisms.

Table 1 illustrates the effects of temperature on key cell performance characteristics. While temperature effects are well documented for solar PV cells [39–43], they are not as well understood for TPVs. A complicating factor is that these coefficients depend on the temperature of the emitter. The results at $T_h = 1160°C$ are represented using best-fit linear temperature coefficients (normalized to 20°C), where the temperature coefficient of a given parameter ($G$) is given by $\beta_G = 1/G(20°C) \cdot (G(T_c) - G(20°C))/(T_c - 20°C)$. We observe that the voltage factor decrease with increasing $T_c$ has the largest effect on efficiency ($\beta_{VF} = -0.326\%/K$). In contrast, spectral management ($SE \cdot IQE$) has a slightly positive coefficient with increasing cell temperature, which is attributed to increasing photocurrent density due to bandgap narrowing. Furthermore, a comparison of the temperature coefficients for the three systems reveals their respective strengths and weaknesses. The spectral efficiency of the planar cell benefits most from increased cell temperature (and its associated bandgap narrowing) due to its relatively high OB absorption. Meanwhile, carrier management in ABC* is least sensitive to increases in cell temperature because of its relatively high $V_{oc}$.

**Table 1 | Temperature coefficients of the ABC, ABC*, and planar cell.**

| Cell | $SE \cdot IQE$ (%/K) | $VF$ (%/K) | $FF$ (%/K) | $\eta$ (%/K) |
|---|---|---|---|---|
| ABC | 0.012 | -0.326 | -0.177 | -0.462 |
| Planar | 0.152 | -0.317 | -0.179 | -0.396 |
| ABC* | 0.012 | -0.188 | -0.11 | -0.277 |

Figure 5a illustrates the heat load ($Q_c$) required to maintain the three cells at a fixed temperature of 20°C, where $Q_c$ is given by: $Q_c = Q_{abs} - P_{mpp} = Q_{abs}(1 - \eta)$. The ABC has a lower heat load than the planar cell owing to its relatively low OB absorption and long carrier lifetime. The longer lifetime has two effects, it decreases the amount of (1) heat generated through carrier recombination, and (2) Joule heating for a given power output because of its relatively high voltage.

With the temperature dependence and heat load described above, we design an active cooling system in which a portion of generated power is diverted and used to circulate the coolant. The temperature of the cell and the emitter are coupled through the heat load and the effective heat transfer coefficient between the cell and coolant. For thermal management, we consider an array of parallel channels embedded in a metal block that supports a 10 cm x 10 cm cell array as shown in Fig. 5b. An internal convection model is used to describe the heat transfer and estimate the power requirements for running a fan to circulate air through the channels (see Appendix A). Figure 5c shows the optimal cell temperature as a function of emitter temperature. For example, an ABC* cell at 40°C maximizes the efficiency for an emitter temperature of 1027°C. In this scenario, 2.8% of the cell output power is consumed by the fan. From this analysis, we observe that the optimal cell temperatures of the ABC are generally lower compared to ABC* and higher than that of the planar design.

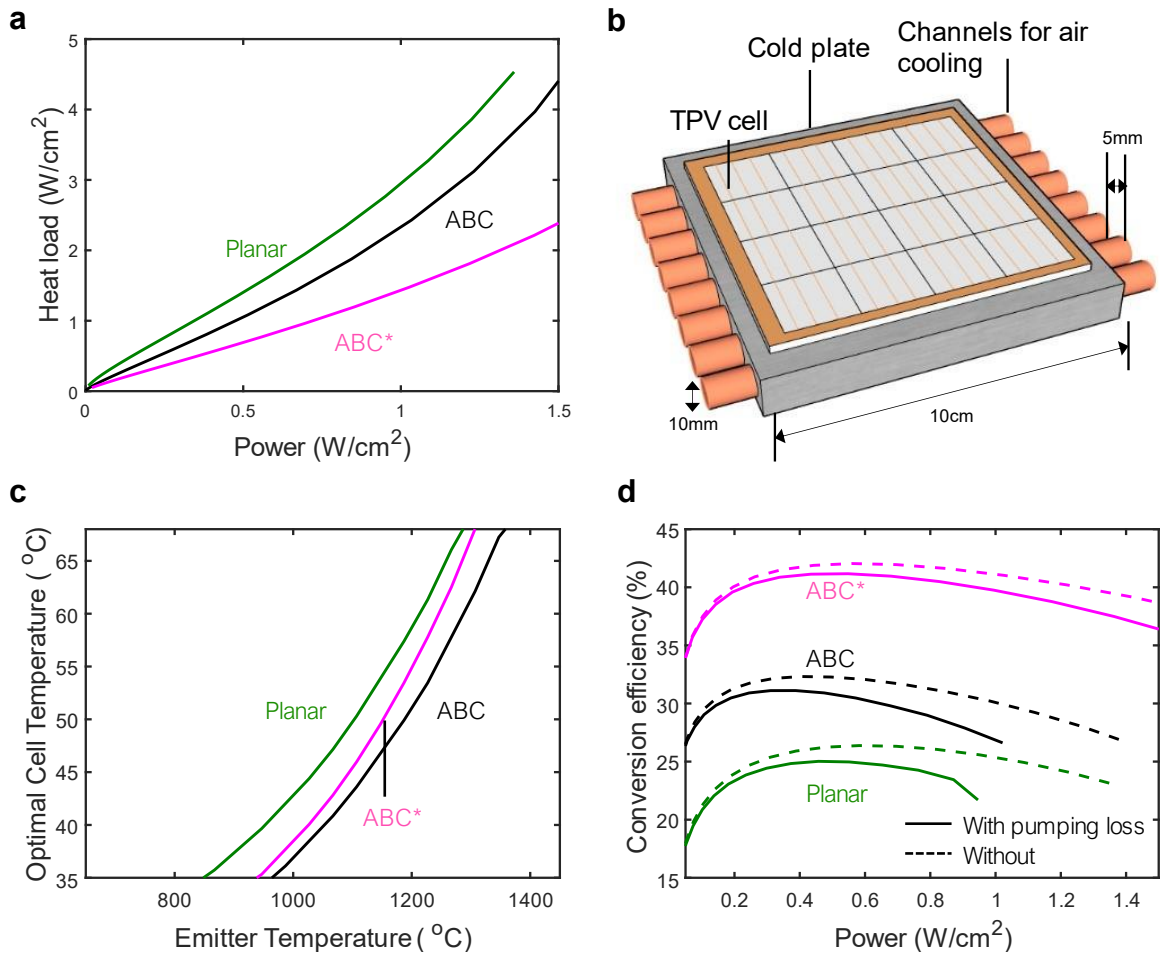

**FIG. 5 | Thermal management of ABCs. (a)** Heat load as a function of power produced for the three configurations. **(b)** Schematic showing envisioned air-cooled thermal management design. **(c)** Optimal cell temperatures as a function of emitter temperature. **(d)** Efficiency as a function of power produced when air-cooling is considered.

Figure 5d shows how cell heating and cooling power affect the efficiency of the overall system. The dashed lines show the power produced at a constant cell temperature of 20°C without accounting for cooling (same as Fig. 4a). The solid lines show the net power produced after diverting some of the generated power to keep the cells at the optimal temperatures in Fig. 4c. Notably, thermal effects are significant at high power densities in all three systems. Furthermore, the optimal operating conditions generally shift toward lower power densities when cooling is considered. For example, before cooling, ABC and planar have optimal efficiencies at emitter temperatures of 1107°C and 1187°C, respectively. After considering cooling, these temperatures are reduced to 1067°C and 1147°C respectively.

Despite the thermal penalties, the air-cooled ABC* system is still predicted to achieve a high-power density (0.5 W/cm$^2$) with >40% efficiency. Beyond air-cooled parallel channels, further improvements to the thermal management design are likely to enable better performance. These include liquid coolants, phase-change heat transfer [44–46], nanostructured surfaces [47,48], and microchannel heat sinks [49,50].

## IV. CONCLUSIONS

This work provides a deeper understanding of how losses in recently demonstrated InGaAs airbridge cells (ABCs) depend on emitter and cell temperature, which is relevant for use of ABCs in practical energy systems at elevated power densities. We find that the efficiency is primarily limited by defect-mediated recombination and series resistance. Realizing improved material quality in InGaAs ABCs, addresses both challenges and enables better performance at higher power densities, including efficiencies exceeding 40% at 0.5 W/m$^2$. Notably, this efficiency takes into consideration the power needed to cool the cells. The work provides additional design guidelines for improving TPV performance including how the optimal emitter temperature decreases when thermal management is considered. For the InGaAs ABC, the optimal emitter temperature is around 1000ºC which is a relatively low temperature for TPV applications. This highlights a promising feature of the ABC design, which is that lowering the emitter temperature reduces the amount of power needed to cool the cells without incurring large spectral management penalties.

**ACKNOWLEDGMENTS**

This work is supported by the National Science Foundation under Grant No. 2038441 and Grant No. 2018572, and by the Army Research Office (ARO) under award number W911NF-19-1-0279. S.M. acknowledges support from the National Science Foundation Graduate Research Fellowship Program under Grant No. DGE-1256260.

[31] J. Talghader and J. S. Smith, *Thermal Dependence of the Refractive Index of GaAs and AlAs Measured Using Semiconductor Multilayer Optical Cavities*, Appl. Phys. Lett. **335**, 335 (1995).

[32] M. Bertolotti, V. Bogdanov, A. Ferrari, A. Jascow, N. Nazorova, A. Pikhtin, and L. Schirone, *Temperature Dependence of the Refractive Index in Semiconductors*, J. Opt. Soc. Am. B **7**, 918 (1990).

[33] T. Burger, C. Sempere, B. Roy-Layinde, and A. Lenert, *Present Efficiencies and Future Opportunities in Thermophotovoltaics*, Joule.

[34] W. Shockley and H. J. Queisser, *Detailed Balance Limit of Efficiency of P-n Junction Solar Cells*, J. Appl. Phys. **32**, 510 (1961).

[35] T. Burger, C. Sempere, and A. Lenert, *Thermophotovoltaic Energy Conversion: Materials and Device Engineering*, in *Nanoscale Energy Transport* (2020), pp. 17-1-17–26.

[36] O. Dupré, R. Vaillon, M. A. Green, O. Dupré, R. Vaillon, and M. A. Green, *A Thermal Model for the Design of Photovoltaic Devices*, in *Thermal Behavior of Photovoltaic Devices* (Springer International Publishing, 2017), pp. 75–103.

[37] F. Kersten, P. Engelhart, H.-C. Ploigt, A. Stekolnikov, T. Lindner, F. Stenzel, M. Bartzsch, A. Szpeth, K. Petter, J. Heitmann, and J. W. Müller, *Degradation of Multicrystalline Silicon Solar Cells and Modules after Illumination at Elevated Temperature*, Sol. Energy Mater. Sol. Cells **142**, 83 (2015).

[38] I. Kaaya, J. Ascencio-Vásquez, K.-A. Weiss, and M. Topič, *Assessment of Uncertainties and Variations in PV Modules Degradation Rates and Lifetime Predictions Using Physical Models*, Sol. Energy **218**, 354 (2021).

[39] O. Dupré, R. Vaillon, and M. A. Green, *Physics of the Temperature Coefficients of Solar Cells*, Sol. Energy Mater. Sol. Cells **140**, 92 (2015).

[40] G. Makrides, B. Zinsser, G. E. Georghiou, M. Schubert, and J. H. Werner, *Temperature Behaviour of Different Photovoltaic Systems Installed in Cyprus and Germany*, Sol. Energy Mater. Sol. Cells **93**, 1095 (2009).
19

## Supplementary information

**Appendix A**: Air-cooled active thermal management of TPV cells

Heat dissipated in the cell ($Q_c$) is removed by forced air, according to:

$$Q_c * A_{array} = h \cdot n_{ch} \cdot \pi \cdot D \cdot L \cdot (T_{wall} - T_{air})$$

where $A_{array}$ is the area of the cell array (10 cm x 10 cm), $h$ is the internal convection heat transfer coefficient, $L$ is the length of channel (10 cm), $D$ is the diameter of channel, $n_{ch}$ is the number of channels.

The heat transfer coefficient is given by [51]:

$$h = \frac{Nu \cdot k_{air}}{D_H}$$

where $Nu$ is the Nusselt number, $D_H$ is the hydraulic diameter, and $k_{air}$ is the thermal conductivity of air. The Nusselt number is calculated from the Reynolds number $Re$ and the Prandtl number $Pr$ (which is 0.7 for air), according to:

$$Nu = 0.023 \cdot Re^{0.8} Pr^{0.4}$$

$$Re = \frac{v \cdot D_H}{\mu}$$

The pumping power, $P_{pump}$, is given by the following set of equations:

$$P_{pump} = \frac{m \cdot \Delta P}{\rho \cdot \eta_{fan} \cdot \eta_{motor}}$$

$$\Delta P = \rho \cdot g \cdot H$$

$$H = \frac{v^2 \cdot f}{2g}\left(\frac{L}{D_H}\right)$$

$$f = 0.316 \cdot Re^{-0.25}$$

where $\Delta P$ is the pressure drop, and the product of the fan efficiency $\eta_{fan}$ and motor efficiency $\eta_{motor}$ was assumed to be 0.8 [51].

From this model, the internal convection resistance of the air-cooled channels represents ~98% of the thermal resistance. Thus, the conductive resistance through the thin cell, including the air gap, and the copper heat sink are negligible.